\documentclass[10pt,A4paper]{article}

\usepackage{arxiv}
\usepackage{url}

\providecommand{\vec}[2][ ]{\left[ #2 \right]^{#1}}

\newcommand\norm[1]{\left\lVert#1\right\rVert}

\usepackage{tikz}
\usetikzlibrary{positioning}
\usetikzlibrary{decorations.pathmorphing,patterns}
\usetikzlibrary{calc,patterns,decorations.markings}
\usetikzlibrary{shapes,arrows}

\usepackage{pgfplots}
\usepackage{algorithm,algpseudocode}
\usepackage{algorithmicx}
\usepackage{authblk}
\usepackage{amsmath}
\usepackage{amsfonts}
\usepackage{gensymb}
\usepackage{siunitx}
\pgfplotsset{every axis plot/.style={thick}}

\DeclareMathOperator*{\argmin}{arg\,min}

\title{PGD-based advanced nonlinear multiparametric regressions for constructing metamodels at the scarce-data limit}

\author[1,2,3]{Abel Sancarlos}
\author[1]{Victor Champaney}
\author[2]{Jean-Louis Duval}
\author[3]{El\'ias Cueto}
\author[1,2]{Francisco Chinesta}

\affil[1]{{\small PIMM Lab. ENSAM Institute of Technology. Paris, France.}}
\affil[2]{{\small ESI Group chair. ENSAM Institute of Technology. Paris, France.}}
\affil[3]{{\small Aragon Institute of Engineering Research. Universidad de Zaragoza. Zaragoza, Spain.}}

\begin{document}

\maketitle

\begin{abstract}
Regressions created from experimental or simulated data enable the construction of metamodels, widely used in a variety of engineering applications. Many engineering problems involve multi-parametric physics whose  corresponding multi-parametric solutions can be viewed as a sort of computational vademecum that, once computed offline, can be then used in a variety of real-time engineering applications including optimization, inverse analysis, uncertainty propagation or simulation based control. Sometimes, these multi-parametric problems can be solved by using advanced model order reduction ---MOR--- techniques. However, when the solution of these multi-parametric problems becomes cumbersome, one possibility consists in solving the problem for a sample of the parametric values, and then creating a regression from all the computed solutions, to finally infer the solution for any choice of the problem parameters. However, addressing high-dimensionality at the low data limit, ensuring accuracy and avoiding overfitting constitutes a difficult challenge. The present paper aims at proposing and discussing different PGD-based advanced regressions enabling the just referred features.\end{abstract}

\section{Introduction}
\label{intro}

Model Order Reduction ---MOR--- techniques express the solution of a given problem (expressed as a partial differential equation ---PDE---, for instance) into a reduced basis with strong physical or mathematical content. Very often, these bases are extracted from solutions of the problem at hand obtained offline. This can be done, for instance, by invoking the proper orthogonal decomposition ---POD--- or the reduced basis method ---RB---\cite{MOR_Paco}. When computing with a reduced basis, the solution complexity scales with the size of this basis, which is in general much smaller than the size of the multi-purpose approximation basis associated with the finite element method ---FEM---, whose size scales with the number of nodes in the mesh. 

Even if the use of a reduced basis implies a certain loss of generality, it enables impressive computing time savings and, as soon as the problem solution continues living in the space spanned by the reduced basis, the computed solution remains accurate enough. Obviously, as soon as one is interested in a solution that can not be accurately approximated within the space spanned by that reduced basis, the solution will be computed fast, but its accuracy is expected to be poor.  To improve generality while ensuring accuracy, an appealing route consists of constructing the reduced basis and solving the problem simultaneously, as the Proper Generalized Decomposition ---PGD--- does \cite{MOR_Paco}. However, this option becomes in general very intrusive, even more than the ones based on the employ of reduced bases. 

To alleviate intrusiveness, non-intrusive procedures were proposed. They proceed by constructing the parametric solution of the parametric problem from a number of high-fidelity solutions performed offline. In general, these are very expensive from the computing time viewpoint, for different choices of the model parameters that constitutes the design of experiments ---DoE---. 

Among these techniques we can mention standard polynomial approximations on sparsely sampled parametric domains. Despite its simplicity, its use is not to be taken lightly. The use of orthogonal polynomial bases, with their associated Gauss-Lobatto points as DoE, allows us to obtain very accurate approximations. However, the sampling (DoE) increases exponentially with either the number of dimensions of the considered polynomial degree. Using randomly sampled DoE, or considering an approximation too rich with respect to the available amount of data (underdetermined approximation problem), results in noticeable overfitting effects. A way of attenuating these unfavorable effects, consists in using an approximation basis avoiding over-oscillating phenomena, as kriging approximations, for instance perform successfully \cite{KRI}, being a major protagonist of the so-called surrogate models (or metamodels) \cite{SUR1,SUR2}. Another possibility consists in restricting polynomial approximations to a low degree, e.g., linear or moderately nonlinear regressions.

Other tentatives concern  the proper orthogonal decomposition with interpolation ---PODI---\cite{LY01}, where usual regressions for expressing the dependence of the modal coefficients on the parameters are employed. Within the PGD rationale, Sparse Subspace Learning ---SSL---\cite{SSL} interpolates the pre-computed solutions related to the DoE associated to an structured grid  (Gauss-Lobatto points) over the whole parametric space, by considering a hierarchical approximation basis for interpolating the precomputed solutions. This ensures the separated representation of the interpolated parametric solution. A sparsely sampled counterpart, the so-called sparse PGD, s-PGD, was proposed in \cite{sPGD_Ruben}.

The main limitations of SSL-based regression procedures is the volume of data, which increases exponentially with the number of parameters involved in the model. Thus, when considering $\mathtt P$ parameters, the lowest approximation level, the so-called {\em 0-level}, which consists in a multi-linear approximation (the product of a linear approximation along each parametric dimension), needs $2^\mathtt P$ data (each datum coming in fact from a high fidelity solution). On the other hand, s-PGD reduces the amount of required data, by considering a sparse sampling. However, the fact of combining higher degree approximations (induced by the separated representations) with very reduced amount of data, exacerbates the risk of overfitting. To avoid overfitting, in \cite{sPGD_Ruben} the authors proposed the use of adaptive approximation bases, the so-called Modal adaptive Strategy ---MAS---, whose degree is kept to a minimum in the first PGD modes (first terms of the finite sum decomposition expressing the variables separation which is at the heart of the PGD). This degree is then increased progressively for the calculation of higher level modes. Other choices of the approximation bases were also considered for limiting these spurious over-oscillating behaviors, as for example the employ of kriging. The s-PGD can thus be viewed as a nonlinear regression that makes use of the separation of variables. This enables its use in multi-parametric settings. 

Regressions are widely employed in artificial intelligence in general, and more particularly in supervised scientific machine learning \cite{udrescu2020ai,brunton2016discovering,HERNANDEZ2021109950}, in the development of cognitive or hybrid digital twins \cite{moya2020digital,sancarlos2020rom,chinesta2020virtual} or even in the field of neuroscience \cite{shiffrin2020brain}. Regression can thus be seen as the main ingredient in the automatic construction of models of the surrounding physical reality. This is of utmost importance in the construction of an artificial intelligence able to maneuver in the physical world \cite{moya2020physically,moya2019learning}.

The main issues related to the implementation of regression in the low-data limit concern nonlinear behaviors in multi-parametric settings. This last factor leads to the so-called curse of dimensionality, i.e., the exponential growth in the number of degrees of freedom (equivalently, the number of necessary sampling points in the phase space) that is necessary to obtain accurate results \cite{laughlin2000cover}.

When constructing models, it is always important to keep them as simple as possible. In other words, parsimonious models are always preferable to more complex ones. This principle, known as Occam's razor \cite{udrescu2020ai,brunton2016discovering}, implies that simpler explanations should be preferred among all the available ones to explain any physical phenomenon. In the literature this is achieved by imposing sparsity in the regression \cite{ibanez2019some,ibanez2018multidimensional,hernandez2020deep,brunton2016discovering}. To obtain parsimonious models able to address sparsity, it is thus convenient to perform regression by combining L2 and L1 norms.

This paper aims at proposing robust, general, frugal and accurate regression methodologies able to operate in separated representation settings. For that purpose, three techniques will be proposed and analyzed. The first is based on an Elastic Net regularized formulation, called $rs$-PGD, combining Ridge and Lasso regressions, that make use, respectively, of the L2 and L1 norms. Both use a rich approximation basis and, to avoid overfitting, the former favors specific solutions with smaller coefficients, while the last enforces the sparsest possible solution by retaining those contributing the most to the solution approximation. 

Then, the doubly sparse regression, the so-called $s^2$-PGD technique will be introduced. The last makes use of the Lasso regularization (the one introduced above that looks for the sparsest approximation through the use of the L1-norm) while searching for the sparsest dimensions. 

The third and last technique, the ANOVA-PGD, aims at allying orthogonal hierarchical bases with a more favorable scaling (with respect to the SSL) of the amount of data with the approximation richness. For that purpose, separated representations and sparse approximations (eventually regularized)  will be combined for addressing multiple correlation terms. 

\begin{figure}

        \centering
	\includegraphics[angle=90,
				width=0.75\textwidth]{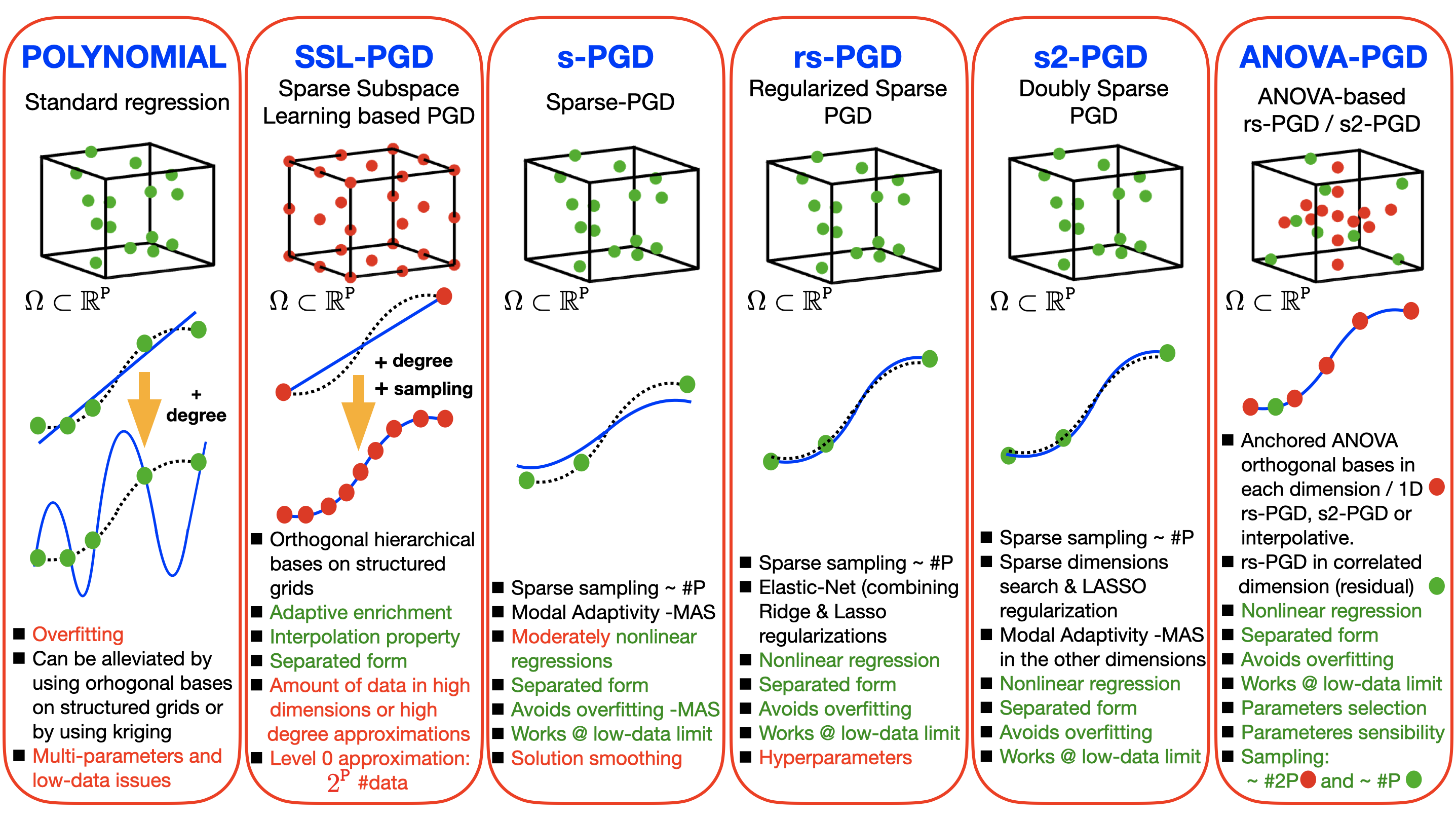}
	\caption{Non-intrusive MOR techniques with the main sampling and approximation features, their pros (emphasized in the green text) and the cons (in red).
	\label{fig1}}
\end{figure}

Figure \ref{fig1} sketches the just referred regression strategies, with the main sampling and approximation features, their pros (emphasized in the green text) and the cons (in red). A comparison on the different exposed techniques, the general workflow for allying them for the solution of a given problem, while addressing their scalability to address industrial problems involving extremely large solutions, constitutes a work in progress that will constitute the part two of the present work.

\section{Regularized regressions: The regularized sparse PGD ($rs$-PGD) and the doubly sparse PGD ($s^2$-PGD)}
\label{reg_s2_PGD}

In this Section, the novel numerical techniques, the regularized sparse PGD ($rs$-PGD) and the doubly-sparse PGD ($s^2$-PGD), are presented and discussed. The content is divided according to the following subsections:
\begin{itemize}
\item In the subsection \ref{$s$-PGD}, the theorical background, from which the proposed methodologies are developed, is presented.
\item In subsection \ref{reg_$s$-PGD}, the regularized PGD is presented starting from the concepts discussed in \ref{$s$-PGD}.
\item In subsection \ref{s$s$-PGD}, the $s^2$-PGD is presented starting from the concepts presented in \ref{reg_$s$-PGD} and \ref{$s$-PGD}.
\end{itemize}

\subsection{Theoretical background: the $s$-PGD}
\label{$s$-PGD}

The $rs$-PGD and the $s^2$-PGD are constructed from the theoretical background of the $s$-PGD in the context of regression problems.\footnote{We would like to stress the fact that the $s$-PGD is based on some of the ideas of the standard Proper Generalized Decomposition (PGD) method for solving PDEs. For this reason, suggest the reader not familiar with the PGD to review previous works in the field such as \cite{BookPGD_Paco,BookPGD_cueto,Abel_PGDTFM}, to name but a few. } In this section, this theoretical basis is reviewed and discussed.

Let us consider an unknown function whose approximation is precisely the objective of this work: 
\begin{equation*}
f(s^1,...,s^{d}): \Omega\subset \mathbb{R}^{d}  \rightarrow \mathbb{R},
\end{equation*}
which depends on $d$ different variables $s^k$, $k=1,\ldots, d$, considered as dimensions of the state space. 

The sparse PGD ($s$-PGD) approach tries to approximate the function $f$ using a low-rank separated (tensor) representation. As in standard PGD procedures, it approximates the function $f$ using a sum of products of one-dimensional functions each one involving one dimension. Each sum is usually called a mode.

This separated form can be expressed as:
\begin{equation} \label{def_PGD}
{f}(s^1,...,s^{d})\approx \tilde{f}^M(s^1,...,s^{d})=\displaystyle{\sum_{\substack{m=1}}^M 
\prod_{{k = 1}}^{d} \psi_m^k(s^k)
},
\end{equation}
where $\tilde{f}^M$ is the approximate, $M$ is the number of modes and $\psi_m^k$ are the one-dimensional function of the mode $m$ and dimension $k$.

In the $s$-PGD context, functions $\psi_m^k$, $m=1, \dots, M$ and $k=1, \dots ,d$ are expressed from standard approximation functions:
\begin{equation}
\psi_m^k(s^k) =
\sum_{j=1}^{D} N_{j,m}^k(s^k) a_{j,m}^k = (\vec{N}_m^k)^\top \vec{a}_m^k,
\end{equation}
where $D$ represents the number of degrees of freedom (nodes) of the chosen approximation. In addition, $ \vec{N}_m^k$ is a column vector with the set of basis functions for the $k$-th dimension and the $m$-th mode and $\vec{a}_m^k$ is a column vector with the coefficients for the $k$-th dimension and the $m$-th mode. The important issue here is to know which set of basis functions are best suited for the problem at hand. For example, a Fourier basis or a polynomial basis can be selected.


In the context of regression problems, the goal is to find an approximation $\tilde{f}^M$, which minimizes the distance (usually related to the L2-norm) to the sought function
\begin{equation}
\label{main_regr}
\tilde{f}^M = \argmin_{f^*}
\displaystyle{\sum_{\substack{i=1}}^{n_t}}
\norm{ f(\vec{s}_i) - {f^*}(\vec{s}_i) 
}_2^2,
\end{equation}
where $\tilde{f}^M$ takes the separated form of Eq. \eqref{def_PGD}, $n_t$ is the number of sampling points to train the model and $\vec{s}_i$ are the different vectors which contain the data points of the training set and the L2-norm is defined from $\norm{f(\vec{s}_i) - {f^*}(\vec{s}_i)}_2^2 = ( f(\vec{s}_i) - {f^*}(\vec{s}_i) )^2$.

The determination of the coefficients of each one-dimensional function for each mode $m = 1, \ldots, M$ is done by employing a greedy algorithm (described in the next sections) such that, once the approximation up to order $M-1$ is known, the new $M$-th  order term is found using a non-linear solver (Picard or Newton, for instance):
\begin{equation} \label{greedy}
\tilde{f}^M=\displaystyle{\sum_{\substack{m=1}}^{M-1} 
\prod_{{k = 1}}^{n_d} \psi_m^k(s^k)
+
\prod_{{k = 1}}^{n_d} \psi_M^k(s^k).
}
\end{equation}

The final goal of the method is that the function $\tilde{f}$ has to approximate $f$ not only when evaluated in the training set but, notably,  in other previously unseen sampling points. This objective is essentially a particular form of {\em machine learning}. This second goal is more difficult to achieve, yet is more important because this evaluates the predictive ability of the model $\tilde{f}$, that is, the capacity to provide good predictions when the model is fed with previously unseen data. Achieving this is particularly difficult when confronted with a high-dimensional problem, for which data is nearly always sparse and/or scarce.

Indeed, the regression problem described by Eq. \eqref{main_regr} only guarantees that the minimization is satisfied by the training set, without saying anything at different sampling points. Hence, if there is not an abundance of sampling points in the training set, in the low-data limit, high oscillations may appear out of these measured points because of the increased risk of overfitting. Usually, this is an undesirable effect because it affects the predictive ability of the constructed regression model.

In order to tackle this problem, the $s$-PGD uses the Modal Adaptivity Strategy (MAS) to take advantage of the greedy PGD algorithm. The idea is to minimize spurious oscillations out of the training set by starting the PGD algorithm looking for modes with low degree. When it is observed that the residual decreases slowly or stagnates, higher order approximation functions are introduced. By doing this, oscillations are reduced, since a higher-order basis will try to capture only what remains in the residual.\footnote{We recommend the reading of \cite{sPGD_Ruben} and \cite{sancarlos2020rom} for more information about the MAS.}

The MAS has proved to be a good strategy to improve significantly the $s$-PGD performance in many problems, see for instance \cite{PhD_Ruben,Abel_Motors1,sancarlos2020rom,PhD_Clara}. However, it has some limitations. For example, it has been observed that the desired accuracy is not achieved before reaching overfitting or the algorithm stops too early when using MAS in some cases. This last issue implies a PGD solution composed of low order approximation functions, thus not getting an as rich as desired function.

In addition, in problems where just a few terms of the interpolation basis are present (that is, there are just some sparse non-zero elements in the interpolation basis to be determined), the strategy fails in recognizing the true model and therefore converging to other one whose predictive performances are bad.

To solve these difficulties, the $rs$-PGD and the $s^2$-PGD are proposed in what follows. Specifically, the first one is used to increase the predictive capacity beyond the $s$-PGD capabilities and the second one is used to sparse identification and variable selection to construct parsimonious models with great explanatory and predictive capabilities.

\subsection{$rs$-PGD}
\label{reg_$s$-PGD}

For the ease of the exposition and representation but without loss of generality, let us continue by assuming that the unknown objective function $f(x,y)$ lives in $\mathbb{R}^{2}$,
\begin{equation*}
f(x,y): \Omega\subset \mathbb{R}^{2}  \rightarrow \mathbb{R},
\end{equation*}
and that it is to be recovered from scarce data.

The goal is therefore to find a function $\tilde{f}^M$ which minimizes the distance to the sought function:
\begin{equation*}
\tilde{f}^M = \argmin_{f^*}
\displaystyle{\sum_{\substack{i=1}}^{n_t}}
\norm{ f(x_i,y_i) - {f^*}(x_i,y_i) 
}_2^2,
\end{equation*}
and that takes the separated form
\begin{equation*}
\tilde{f}^M(x,y)=\displaystyle{\sum_{\substack{m=1}}^M 
X_m(x)\cdot Y_m(y)
}
=
\displaystyle{\sum_{\substack{m=1}}^M 
\Big(
(\vec{N}_m^x)^\top \vec{a}_m^x
\cdot
(\vec{N}_m^y)^\top \vec{a}_m^y
\Big)
},
\end{equation*}
where $n_t$ is the number of sampling points employed to train the model (training set). Here, the superscript $M$ is employed to highlight the rank of the sought function. How to determine the precise value of $M$ will be detailed hereafter. 

In the PGD framework, an iterative scheme based on an alternating direction strategy is usually used to solve the resulting non-linear problem ---note that we look for products of one-dimensional functions--- and compute $\vec{a}_M^x$ and $\vec{a}_M^y$. This strategy computes $\vec{a}_M^{x,k}$ from $\vec{a}_M^{y,k-1}$ and $\vec{a}_M^{y,k}$ from $\vec{a}_M^{x,k}$ where $\vec{a}_M^{y,k}$ indicates the values of $\vec{a}_M^{y}$ at iteration $k$ of the nonlinear iteration algorithm.
The iterations proceed until reaching a fixed point according to a user-specified tolerance.

Defining $\vec{N}_m^x(x_i)$ and $\vec{N}_m^y(y_i)$ as the vectors containing the evaluation of the interpolation basis of the $m^{th}$ mode at $x_i$ and $y_i$, respectively, we can write the following matrix equations defining the systems to solve:
\begin{equation}
\label{eqpgd_x}
\mathbf{M}_x
\cdot
\vec{a}_M^x
=
\vec{r},
\end{equation}
\begin{equation}
\label{eqpgd_y}
\mathbf{M}_y
\cdot
\vec{a}_M^y
=
\vec{r},
\end{equation}
where:
\begin{align*}
\vec{r}
&
=
\begin{pmatrix} 
f(x_1,y_1)-\tilde{f}^{M-1}(x_1,y_1) 
\\ 
\vdots 
\\
f(x_{n_t},y_{n_t})-\tilde{f}^{M-1}(x_{n_t},y_{n_t}) 
\end{pmatrix},
\\
\mathbf{M}_x
&
=
\begin{pmatrix} 
(\vec{N}_M^y(y_1))^\top \vec{a}_M^y 
\cdot
(\vec{N}_M^x(x_1))^\top 
\\ 
\vdots 
\\
(\vec{N}_M^y(y_{n_t}))^\top \vec{a}_M^y 
\cdot
(\vec{N}_M^x(x_{n_t}))^\top 
\end{pmatrix},
\\
\mathbf{M}_y
&
=
\begin{pmatrix} 
(\vec{N}_M^x(x_1))^\top \vec{a}_M^x 
\cdot
(\vec{N}_M^y(y_1))^\top 
\\ 
\vdots 
\\
(\vec{N}_M^x(x_{n_t}))^\top \vec{a}_M^x 
\cdot
(\vec{N}_M^y(y_{n_t}))^\top 
\end{pmatrix}.
\end{align*}

If Eq. \eqref{eqpgd_x} and \eqref{eqpgd_y} are solved in the Ordinary Least Squares (OLS) sense:
\begin{equation}
\label{ols_x}
\vec{a}_M^x
=
(
\mathbf{M}_x^\top
\mathbf{M}_x
)^{-1}
\cdot
\mathbf{M}_x^\top
\vec{r},
\end{equation}
\begin{equation}
\label{ols_y}
\vec{a}_M^y
=
(
\mathbf{M}_y^\top
\mathbf{M}_y
)^{-1}
\cdot
\mathbf{M}_y^\top
\vec{r}
\end{equation}
which give us the usual matrix equations in the OLS context.

The $rs$-PGD is based on putting a penalty term when solving \eqref{eqpgd_x} and \eqref{eqpgd_y} with the following objectives:
\begin{itemize}
\item To reduce overfitting.
\item To deal with strong multicollinearity, namely when the OLS regression problem is ill-posed.
\end{itemize}

Note that the overfitting problem can easily arise in the $s$-PGD context when high-order approximations (that separated representations exacerbate) are employed because of the usual unstructured low data regime used to train the model. This issue strongly affects the model's ability to perform on new, unseen sets. Therefore, the idea of using the penalty term consists in improving the model's ability to perform on new samples at the cost of increasing the bias or the error model in the training set for a given set of basis functions.

Different regularizations can be envisaged depending on the properties of the problem such as the Tikhonov regularization or the Elastic Net regularization.

For the sake of simplicity but without loss of generality, we start introducing the ridge regression regularization (a special case of the Tikhonov regularization) that will be generalized later to lead to the Elastic Net redularization. 

For this purpose, we first rewrite Eqs. \eqref{ols_x} and \eqref{ols_y}:
\begin{equation}
\label{ridge1x}
\vec{a}_M^x
=
(
\mathbf{M}_x^\top
\mathbf{M}_x
-
\lambda \mathbf{I}
)^{-1}
\cdot
\mathbf{M}_x^\top
\vec{r}
\end{equation}
\begin{equation}
\label{ridge1y}
\vec{a}_M^y
=
(
\mathbf{M}_y^\top
\mathbf{M}_y
-
\lambda \mathbf{I}
)^{-1}
\cdot
\mathbf{M}_y^\top
\vec{r},
\end{equation}
where $\lambda$ is the penalty factor and $\mathbf{I}$ is the identity matrix. In this case, both dimensions are equally penalized but different penalty factors could be considered depending on the considered dimension.

The regularized problems associated to Eqs. (\ref{ridge1x}) and (\ref{ridge1y}) are:
\begin{equation}
\label{ridge2x}
 \vec{a}_M^x = \argmin_{\vec{a}_M^{x\ast}}
\Big\{
\norm{\vec{r} - \mathbf{M}_x \vec{a}_M^{x\ast}}_2^2
+
\lambda
\norm{\vec{a}_M^{x\ast}}_2^2
\Big\},
\end{equation}
\begin{equation}
\label{ridge2y}
 \vec{a}_M^y = \argmin_{\vec{a}_M^{y\ast}}
\Big\{
\norm{\vec{r} - \mathbf{M}_y \vec{a}_M^{y\ast}}_2^2
+
\lambda
\norm{\vec{a}_M^{y\ast}}_2^2
\Big\},
\end{equation}
where the problem is divided in solving a ridge regression problem for each dimension when computing $\vec{a}_M^x$ and $\vec{a}_M^y$ during the alternate direction fixed point strategy.

The interpretation of employing Eqs. \eqref{ridge2x} and \eqref{ridge2y} during the PGD iterative scheme can be thought of as an attempt of solving the following problem within the PGD rationale:
\begin{equation}\label{ridge1xy}
\tilde{f}^M (\vec{a}_M^x,\vec{a}_M^y) 
\\= \argmin_{\vec{a}_M^{x\ast}, \vec{a}_M^{y\ast}}
\Big\{
\norm{f - \tilde{f}^{M}(\vec{a}_M^{x\ast}, \vec{a}_M^{y\ast}) }_2^2
+
\lambda
\norm{\vec{a}_M^{x\ast}}_2^2
+
\lambda
\norm{\vec{a}_M^{y\ast}}_2^2
\Big\},
\end{equation}
where $\norm{ \cdot }_2$ is the Euclidean norm, and $\tilde{f}^M$ is the function defined in \eqref{greedy} where the new $M$-th  order term of the model is sought.

As the terminology used in this section shows, a regularization problem is formulated at each enrichment step. Thus, we are looking for the best penalty factor at each updating stage, adapting the regularization whenever the approach is enriched. Other possibilities can be envisaged but this one seems the one which offers the best results according to our numerical experiments.

A null intercept term was assumed for $\vec{a}_M^x$ and $\vec{a}_M^y$ in the deduction of equations \eqref{ridge1x}, \eqref{ridge1y}, \eqref{ridge2x} and \eqref{ridge2y}. If this term is going to be included, it can be treated as in standard ridge procedures when solving the corresponding linear regularized regression problem for each dimension during the alternating direction strategy.


As we are generally looking for the mode with best predictive abilities in each enrichment, the proposed criterion to choose $\lambda$ is to perform a $k$-fold cross-validation and select the value of $\lambda$ that minimizes the cross-validated sum of squared residuals (or some other measure). It is also possible to use the “one-standard error” rule (heuristic) with cross-validation, in which we choose the most penalized model whose error is no more than one standard error above the error of the best model. Such a rule acknowledges the fact that the tradeoff curve is estimated with error, and hence takes a conservative approach \cite{Key_Stats}.

If enough data is available, the split of the training set in two subgroups is equally a reasonable option to select $\lambda$ and in addition, computationally less demanding. In this case, one subgroup is employed for constructing the model and the other one to evaluate the predictive ability and then to select $\lambda$ accordingly.

The Elastic Net regularization results of including a L1-norm regularization, from which Eqs. (\ref{ridge2x})-(\ref{ridge2y}) and Eq. (\ref{ridge1xy}) become:
\begin{equation}
\label{ENx}
 \vec{a}_M^x = \argmin_{\vec{a}_M^{x\ast}}
\Big\{
\norm{\vec{r} - \mathbf{M}_x \vec{a}_M^{x\ast}}_2^2
+
\lambda
\left[
(1-\alpha)
\norm{\vec{a}_M^{x\ast}}_2^2
+
\alpha   
\norm{\vec{a}_M^{x\ast}}_1
\right]
\Big\},
\end{equation}
\begin{equation}
\label{ENy}
 \vec{a}_M^y = \argmin_{\vec{a}_M^{y\ast}}
\Big\{
\norm{\vec{r} - \mathbf{M}_y \vec{a}_M^{y\ast}}_2^2
+
\lambda
\left[
(1-\alpha)
\norm{\vec{a}_M^{y\ast}}_2^2
+
\alpha   
\norm{\vec{a}_M^{y\ast}}_1
\right]
\Big\},
\end{equation}
and
\begin{multline}\label{EN}
\tilde{f}^M (\vec{a}_M^x,\vec{a}_M^y) = \argmin_{\vec{a}_M^{x\ast}, \vec{a}_M^{y\ast}}
\left \{
\norm{f - \tilde{f}^{M}(\vec{a}_M^{x\ast}, \vec{a}_M^{y\ast}) }_2^2 \right .
\\ 
+ \quad
\lambda
\left[
(1-\alpha)
\left(
\norm{\vec{a}_M^{x\ast}}_2^2
+
\norm{\vec{a}_M^{y\ast}}_2^2
\right)
+
\alpha   
\left(
\norm{\vec{a}_M^{x\ast}}_1
+
\norm{\vec{a}_M^{y\ast}}_1
\right)
\right]
\Big\},
\end{multline}
respectively, where $\alpha \in [0,1)$ and $\lambda$ are the penalty factors. These coefficients could be also different for the different dimensions, and the lambda coefficients also different for the norm L2 and L1. The limit cases $\alpha = 0$ and $\alpha=1$ result in the Ridge and Lasso regressions respectively.

\subsection{$s^2$-PGD}
\label{s$s$-PGD}

For the ease of the exposition and representation but without loss of generality,
let us continue by assuming the same two-dimensional unknown function discussed in Section \ref{reg_$s$-PGD}.

Here, we are dealing with a solution which admits a sparse solution for a certain basis using the PGD separated form \eqref{def_PGD}.
In this case, the goal is to identify the correct non-zero coefficients at each enrichment step in order to guide the approach to the correct separated representation.

Without a roadmap to select these nonzero coefficients, the traditional $s$-PGD fails to capture the true relationship between the model's features as well as its final response. Furthermore, if high-order terms appear in the searched function, this issues become even worse leading to serious overfitting issues.

Let us consider the theory discussed in the previous section but now considering the L1 regularization with the idea to promote sparsity in the overall solution of the nonlinear regression problem:
\begin{equation}\label{ridgexy}
\tilde{f}^M (\vec{a}_M^x,\vec{a}_M^y) 
\\= 
\argmin_{\vec{a}_M^{x\ast}, \vec{a}_M^{y\ast}}
\Big\{
\norm{f - \tilde{f}^{M}(\vec{a}_M^{x\ast}, \vec{a}_M^{y\ast}) }^2_2
+
\lambda
\norm{\vec{a}_M^{x\ast}}_1
+
\lambda
\norm{\vec{a}_M^{y\ast}}_1
\Big\}.
\end{equation}

%

This formulation is convenient because the nonlinear problem can be solved using the PGD constructor, with an alternate direction fixed point strategy, where just a LASSO regression problem is considered in each dimension. 

Therefore, the regression problems for the iterative scheme will be:

\begin{equation}
\label{lassox}
 \vec{a}_M^x = \argmin_{\vec{a}_M^{x\ast}}
\Big\{
\norm{\vec{r} - \mathbf{M}_x \vec{a}_M^{x\ast}}^2_2
+
\lambda
\norm{\vec{a}_M^{x\ast}}_1
\Big\},
\end{equation}
\begin{equation}
\label{rlassoy}
 \vec{a}_M^y = \argmin_{\vec{a}_M^{y\ast}}
\Big\{
\norm{\vec{r} - \mathbf{M}_y \vec{a}_M^{y\ast}}^2_2
+
\lambda
\norm{\vec{a}_M^{y\ast}}_1
\Big\},
\end{equation}
that consists of solving a LASSO regression problem for each dimension when computing $\vec{a}_M^x$ and $\vec{a}_M^y$ within the alternate direction fixed point strategy. Moreover, as previously discussed, in the present case again, both dimensions are equally penalized but different penalty factors could be envisaged.

As we are iteratively solving a LASSO problem in each direction, we will end up with sparse solutions for each one-dimensional function choosing the right penalty factor. Again, a null itercept term was assumed.

In case of looking for sparsity just in the $x$ dimension, only Eq. (\ref{lassox}) applies for computing coefficients $ \vec{a}_M^x$, whereas coefficients $ \vec{a}_M^y$ are calculated by invoking the standard $s$-PGD or the $rs$-PGD, addressed in the previous section.



%
To determine $\lambda$, we first refer the reader to the discussion of the previous section. Then, the following considerations applied in the case of the doubly sparse PGD:
\begin{itemize}
\item Before selecting a model according to the predictive criterion, a filter is considered taking only the models with a minimum sparsity criterion $\norm{\vec{a}_M^x}_0 \leq \chi^{lim}_x$. If sparsity is also desired in $y$ direction, $\chi^{lim}_y$ will be defined accordingly. Note: We define $\norm{\cdot}_0$ by $\norm{\vec{x}}_0 = \#\{ i:\vec{x}_i \neq 0 \}$. We consider this notation even if it is actually not a norm.
\item Once model selection is performed, the OLS methodology is employed with the detected non-zero elements to obtain the correct update. The reason of this step is that LASSO regression terms are in general not accurate, and so it may be necessary to de-bias the obtained values. Remember that the LASSO shrinkage causes the estimates of the non-zero coefficients to be biased towards zero and in general they are not consistent \cite{Kutz_data_driven_2019} \cite{Key_Stats}.
\end{itemize}


\section{The ANOVA-based sparse-PGD}

The ANOVA decomposition of a function $f(s^1, \dots , s^d): \Omega \subset \mathbb{R}^d\rightarrow\mathbb{R}$ is an orthogonal decomposition based on the analysis of variance, a statistical model designed for data analysis. Thus, the function $f(\vec{s})$ can be written as a sum of orthogonal functions:
\begin{equation}
f(\vec{s}) = f_0 + \sum_{i=1}^d f_i(s^i) + \sum_{i_1=1}^d \sum_{i_2 = i_1}^d f_{i_1,i_2} (s^{i_1},s^{i_2})+ \ldots + f_{1,2,\dots d}(s^1,s^2,\dots,s^d),
\end{equation}
satisfying
\begin{equation}
\mathbb E_i(f_{i_1, \dots , i_k}(s^{i_1}, \dots , x^{i_k})) =0,
\end{equation}
where $\mathbb E_i$ refers to the expectation with respect to any coordinate $i$ in the set $(i_1, \dots , i_k)$, $1 \leq k \leq d$. This property results in the orthogonality of functions involved in the previous decomposition. 

To prove it, consider for example a simple 2D case with, $\vec{s}=(x,y)$, $f(\vec{s}) \equiv f(x,y)$. Thus, with $\mathbb E_x(f_x(x))=0$, $\mathbb E_x(f_{x,y}(x,y))=0$ and $\mathbb E_y(f_{x,y}(x,y))=0$, we have $\mathbb E_{x,y}(f_{x,y}(x,y) f_x(x)) = \mathbb E_x \{ \mathbb E_y(f_{x,y}(x,y)) \ f_x(x) \}=0$.

The number of function involved in the decomposition (without considering the constant term) is $2^d-1$, and they can be parametrized by the integer $n$, $n=1, \dots , 2^d-1$. The different functions involved in the ANOVA decomposition can be expressed from expectations according to:
\begin{equation}
\left \{ 
\begin{array}{l}
\mathbb{E}(f(\vec{s})) = f_0 \\
\mathbb{E}(f(\vec{s} | s^i)) = f_i(s^i) +f_0 \\
\mathbb{E}(f(\vec{s} | s^i,s^j)) = f_{i,j}(s^i,s^j) + f_i(s^i) + f_j(s^j) + f_0 \\
\vdots 
\end{array}
\right . 
\end{equation}
where $\mathbb{E} (f(\vec{s} | s^i))$ refers to the integration on all the variables except $s^i$. 

\subsection{Sensitivity analysis: Sobol coefficients}

The variance of $f(\vec{s})$, $\mathtt{Var}(f(\vec{s}))$, taking into account the orthogonality of the functions involved in the ANOVA decomposition, reads
\begin{equation}
\mathtt{Var}(f(\vec{s})) = \sum_{n=1}^{2^d-1} \mathbb E\left ( f_n(\vec{s}_n)\right )^2  = \sum_{n=0}^{2^d-1} \mathtt{Var}_n,
\end{equation}
that allows defining the so-called Sobol sensitivity coefficients $\mathcal S_n$
\begin{equation}
\mathcal S_n = \frac{\mathtt{Var}_n}{\mathtt{Var}(f(\vec{s}))}.
\end{equation}

\subsection{The {\em anchored} ANOVA}

Multidimensional settings imply expensive calculations for computing the multidimensional expectations. For alleviating those costly computations we introduce the so-called anchor point $\vec{c}$ such that $f_0 =f(\vec{c})$. Then, in the definition of the functions involved in the ANOVA decomposition, the expectations are replaced by $f(\vec{c} | \vec{s}_n)$, that is, the particularization of the function in the anchor point, except for those coordinates involved in $\vec{s}_n$.

\subsection{Combining the {\em anchored}-ANOVA with the  sparse PGD}

A valuable strategy consists in: (i) first, using the standard {\em anchored}-ANOVA for evaluating the functions depending on each dimension $f_i(s^i)$, $i=1, \dots , d$, by suing an adequate sampling, a sort of multidimensional cross centered at the anchor point $\vec{c}$. In each dimension, $f_i(s^i)$ can be approximated by using any variable approximation, eventually the regularized ones discussed in the previous sections. Then, (ii) one could compute the residual $f^\prime(\vec{s})$:
\begin{equation}
f^\prime(\vec{s}) = f(\vec{s}) - f_0 - \sum_{i=1}^d f_i(s^i),
\end{equation}
and finally, (iii) using the $rs$-PGD, or the$s^2$-PGD, for approximating that residual $f^\prime(\vec{s})$ that contains the different correlations. In that case, an enhanced sparse-sampling can be considered, trying to approach as much as possible the points involved in the sparse sampling to the borders of the parametric domain.

\section{Results}
\label{res_PGD}

In this section, the results of using the above techniques are shown for different cases.
First, in Section \ref{res_regpgd}, the error reduction is shown when using the $rs$-PGD comparing with the classical procedure ($s$-PGD).
Then, in Section \ref{res_s2pgd}, sparse identification and error reduction is presented when using the $s^2$-PGD comparing with the standard sparse procedure ($s$-PGD). Finally, Section \ref{anov} employes the analysis of variance and combines it with regularized approximations to define an original and powerful regression methodology.

\subsection{Results for the $rs$-PGD approach}
\label{res_regpgd}
The following examples considers the Elastic Net Regularization. For that purpose, an $\alpha$ parameter is employed for combining the Ridge and Lasso regression.
The $\alpha$ parameter is selected by running the algorithm several times for different $\alpha$ values, and then choosing the one which has better predictive performances.

\subsubsection{A first example involving a five dimensional polynomial}

In the first example, we are trying to approximate the five-dimensional function
\begin{equation}\label{Ex1}
f(x_1, x_2, x_3, x_4, x_5) = (8x_1^3 - 6x_1 - 0.5x_2)^2
+ (4x_3^3 - 3x_3 - 0.25x_4)^2
+ 0.1(2x_5^2 - 1).
\end{equation}

The above function is intended to be reconstructed in the domain $\Omega = [-0.51, 0.51]^5$.
The sampling for the training set contains 160 points. Therefore, only these points are used to construct the model either using the $s$-PGD or the $rs$-PGD methodology. In addition, the Latin hypercube sampling (LHS) is used to generate this set of data.

On the other hand, a testing set of 54000 untrained points is considered to compare the results between techniques when predicting unseen scenarios. This second set will be used to study the predictive ability of both models once they are finally constructed.

A standard MAS employing up to 4th degree polynomials for both the s-PGD and the $rs$-PGD is considered. 
To measure the error of both methodologies in the testing set, the following error criterion is used:
\begin{align*}
\text{err}_{pgd} &= \frac{ \norm{ \vec{z} - \vec{z}_{pgd} }_2 } { \norm{\vec{z}}_2 }; &
\text{err}_{rpgd} &= \frac{ \norm{ \vec{z} - \vec{z}_{rpgd} }_2 } { \norm{\vec{z}}_2 };
\end{align*}
where $\vec{z}$ is the vector containing the values of $f(x_1, x_2, x_3, x_4, x_5)$ in the testing set, $\vec{z}_{pgd}$ and $\vec{z}_{rpgd}$ are the vectors containing the prediction in the testing set of both methodologies ($s$-PGD and $rs$-PGD, respectively).

After employing the discussed techniques in the above conditions, we obtain in this example that the error is reduced by 52.38 \% using the $rs$-PGD with $\alpha = 0.1$.

\begin{figure}
\begin{center}
\includegraphics[width=\linewidth]{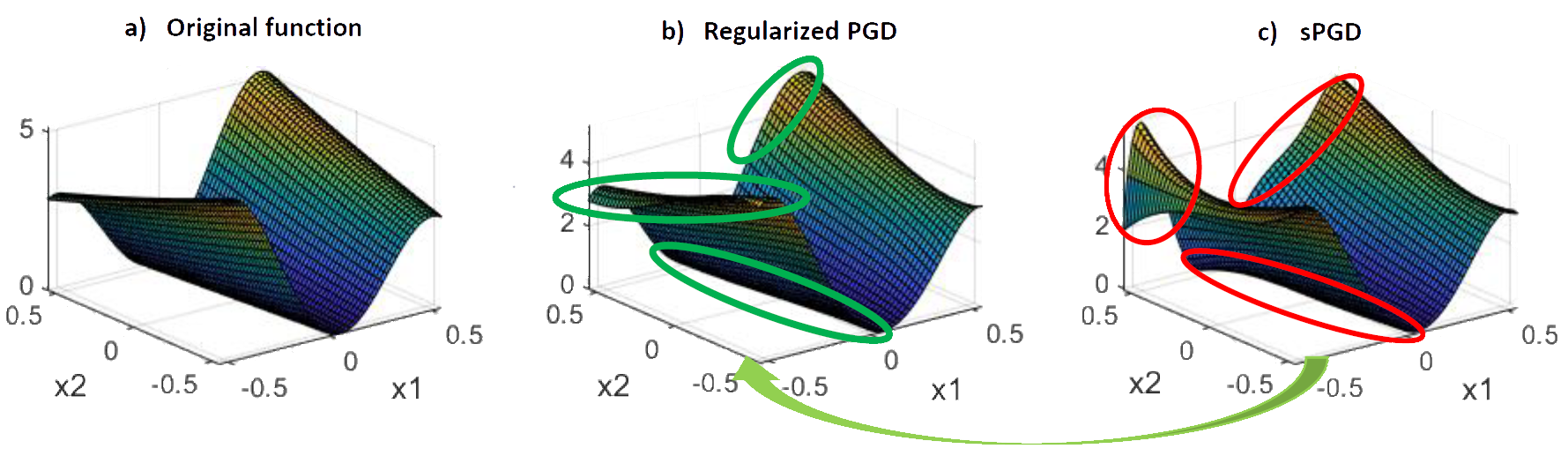}
\caption{Comparing the reference (Eq. \eqref{Ex1}) and its associated $s$-PGD and $rs$-PGD regressions, at points $(x_1,x_2,x_3=0,x_4=0,x_5=0.7071)$}\label{rPGD_ex1a}
\end{center}
\end{figure}

To perceive the improvements and the overfitting reduction, in Figure \ref{rPGD_ex1a}, we show a plot of the original function $f(x_1,x_2,x_3=0,x_4=0,x_5=0.7071)$. It can be noticed  that the $rs$-PGD corrects the shape of the function in the indicated areas in Fig. \ref{rPGD_ex1a}, improving the performance of the regression.

\begin{figure}
\begin{center}
\includegraphics[width=\linewidth]{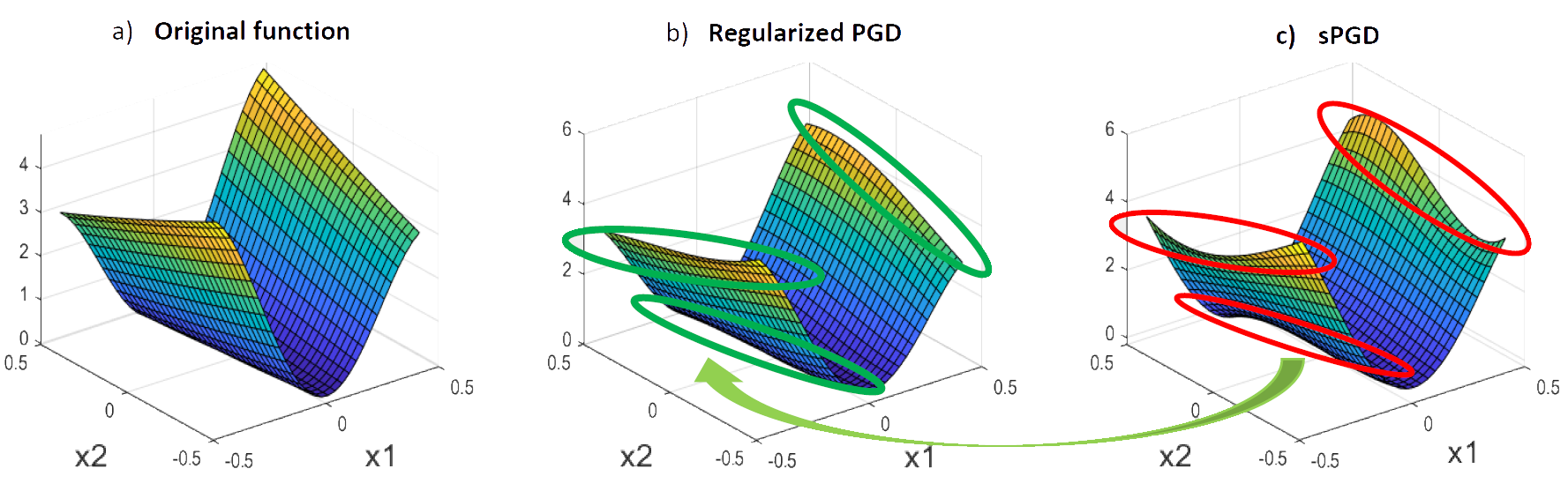}
\caption{Comparing the reference (Eq. \eqref{Ex1}) and its associated $s$-PGD and $rs$-PGD regressions, at points $(x_1,x_2,x_3=-0.17069,x_4=-0.17069,x_5=-0.015517)$}\label{rPGD_ex1b}
\end{center}
\end{figure}

This improvement occurs over the whole five-dimensional domain. Other result is shown in Figure \ref{rPGD_ex1b} that depicts $f(x_1,x_2,x_3=-0.17069,x_4=-0.17069,x_5=-0.015517)$.

\subsubsection{A second example involving five dimensions with trigonometric and logarithmic functions}

In this second example, we are trying to approximate the function:
\begin{equation}\label{Ex2}
\begin{aligned}
f(x_1, x_2, x_3, x_4, x_5) = & \cos(x_1 x_2) 
\Big[
\big( \sin(2x_3) - 3.14 \big) \log(3x_4 + 1.5) \cos(x_5) \\
+ &
\exp(x_4) \cosh(x_3) \sinh(x_5)
\Big],
\end{aligned}
\end{equation}
by using the $rs$-PGD with polynomials.
The above function is intended to be reconstructed in the domain $\Omega = [-1, 1]^5$.

In this case, the sampling for the training set contains 290 points. Therefore, only these points are used to construct the model either by using the $s$-PGD or the $rs$-PGD methodology. In addition, the Latin hypercube sampling is used to generate this set of data.

On the other hand, a testing set of 2000 untrained points is available to compare the results when predicting unseen scenarios.
Again a standard MAS is employed reaching 4th degree polynomials in both, the s-PGD and the $rs$-PGD. 
An error reduction of about 47\% is accomplished with $\alpha = 0.5$.

\subsubsection{Results on the chaotic Lorenz system.}
\label{lorentz}
As a last example, we consider a canonical model for chaotic dynamics, the Lorentz system \cite{Lorentz_1,Lorentz_2}:
\begin{align*}
  \dot{x} &= \sigma (y - x)\\ 
  \dot{y} &= x (\rho - z) - y \\ 
  \dot{z} &= xy - \beta z
\end{align*}
with parameters $\sigma = 10$, $\rho = 28$ and $\beta = 8/3$.

Data are collected using a sampling without replacement in the interval $t \in [0,20]$ until completing a set of 102 points. These 102 points will be divided in two sets: the construction set and the validation set. The first one will be used to compute the regression coefficients and the other one to select the hyperparameters. Furthermore, the ridge regularization is employed as well as the MAS for the $rs$-PGD when identifying the dynamics. As in other instances, the Chebyshev basis is used in the one-dimensional approximations.

The $rs$-PGD successfully detects the important non-zero coefficients with an error below  0.02 \% in the construction and validation set for the three variables. As an illustration, the initial identified coefficients for $\dot{x}$ are shown in Table \ref{coef_xdot}. As we can observe, the theoretical zero coefficients are not always exactly zero but they are very small. However, we are dealing with a chaotic dynamics where very small deviations on the identified parameters can produce huge deviations in the long-time predictions.

\begin{table}[h]
\begin{center}
\begin{tabular}{|c|c|}
\hline
 & $\dot x$ \\ \hline \hline
 $\{ \text{' '} \}$ & $\{[8.7112e-04]\}$ \\ \hline
 $\{ 'x' \}$ & $\{[-9.9997]\}$ \\ \hline
 $\{ 'y' \}$ & $\{[9.9996]\}$ \\ \hline
 $\{ 'z' \}$ & $\{[0]\}$ \\ \hline
 $\{ 'xy' \}$ & $\{[-1.3783e-05]\}$ \\ \hline
 $\{ 'xz' \}$ & $\{[0]\}$ \\ \hline
 $\{ 'yz' \}$ & $\{[0]\}$ \\ \hline
 $\{ 'xyz' \}$ & $\{[0]\}$ \\ \hline
\end{tabular}
\end{center}
\caption{Initial $rs$-PGD model for $\dot{x}$}\label{coef_xdot}
\end{table}%

\begin{figure}
\begin{center}
\includegraphics[width=1\linewidth]{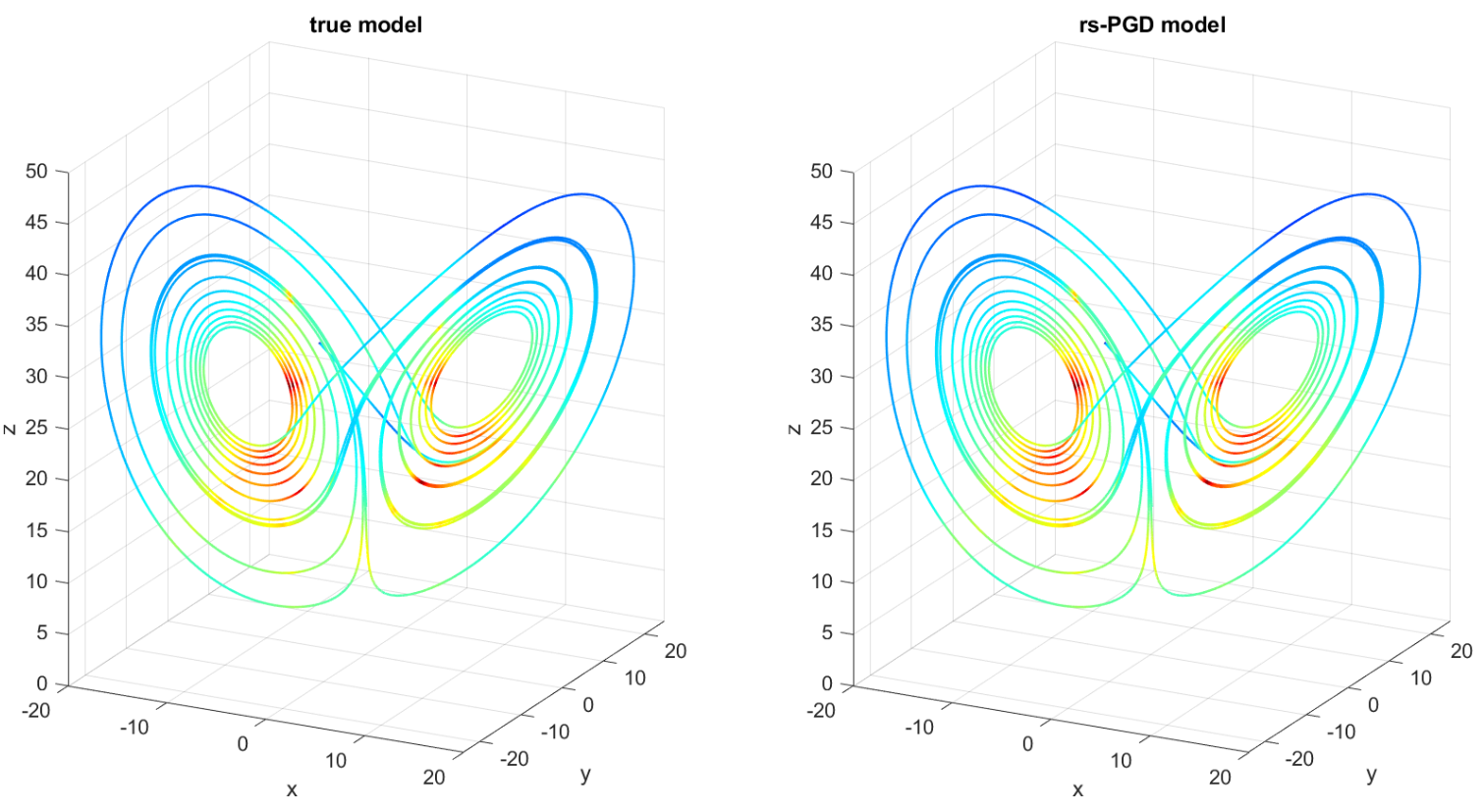}
\caption{True dynamics and dynamics identified by the $rs$-PGD model}\label{lorentz_butterfly}
\end{center}
\end{figure}

\begin{figure}
\begin{center}
\includegraphics[width=1\linewidth]{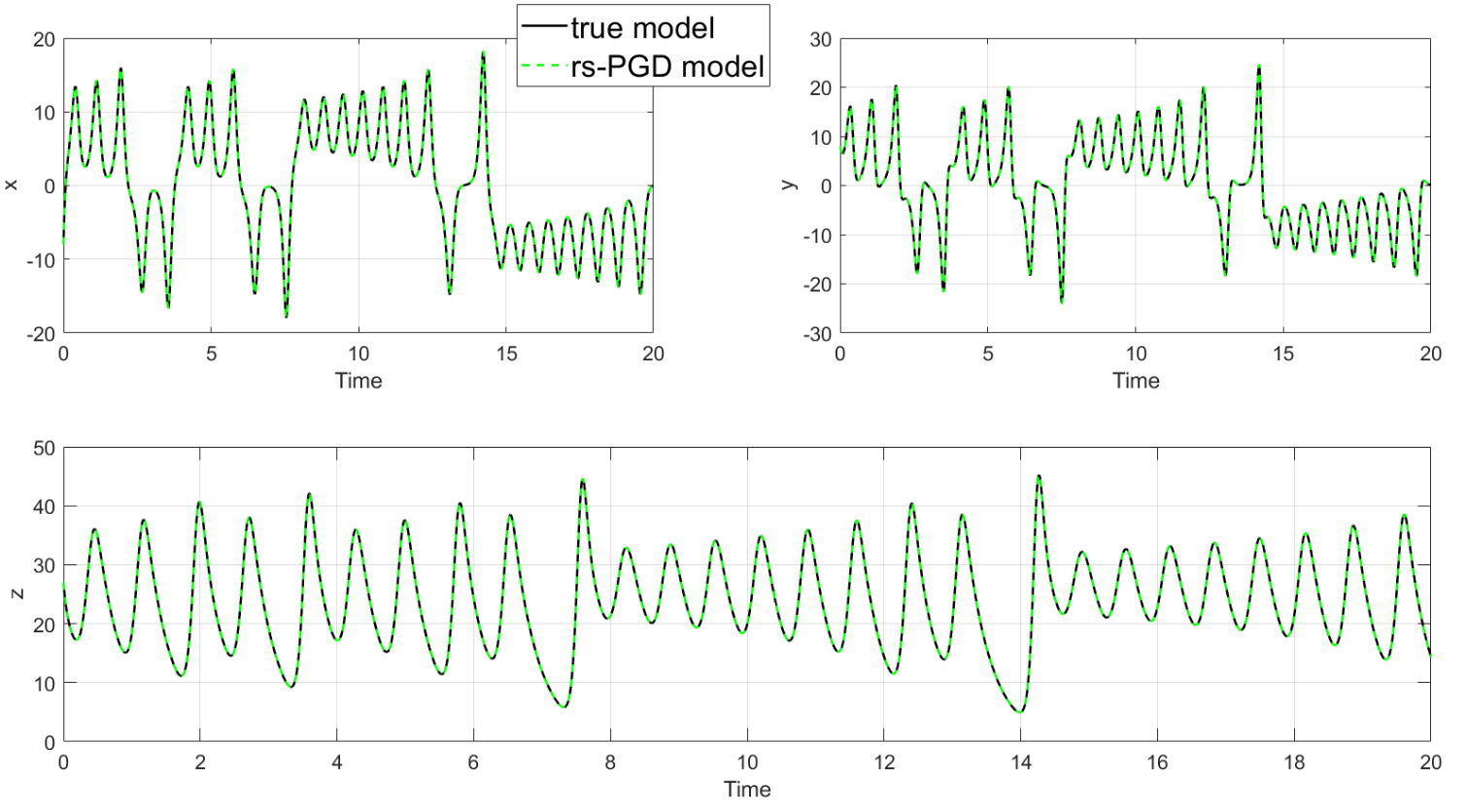}
\caption{Comparison between $rs$-PGD predictions and true dynamics on the three-variable time evolution.}\label{lorentz_xyz}
\end{center}
\end{figure}

For this reason, once the $rs$-PGD solution is computed, a filter based on the sequential thresholded least-squares (STLS) can be applied on the remaining coefficients, aiming at removing the coefficients below a given threshold, and then the least squares procedure applies again to recompute the coefficients, for obtaining a very accurate regression, as Figs. \ref{lorentz_butterfly} and \ref{lorentz_xyz} prove.

\subsection{Checking the performances of $s^2$-PGD when addressing sparse solutions}
\label{res_s2pgd}

\subsubsection{A first example involving sparsity in one dimension}

In the first example of this Section, we are trying to approximate the function:
\begin{equation} \label{eqn:s2PGD_a}
f(x_1, x_2, x_3) = (\sin(2x_1) - 3.14)  T_5(x_2)
+ \exp(x_3) \cosh(x_1),
\end{equation}
by using a Chebyshev basis for the one-dimensional functions of the PGD. The above function is intended to be reconstructed in the domain $\Omega = [-1, 1]^3$.

\begin{figure}
\begin{center}
\includegraphics[width=0.7\linewidth]{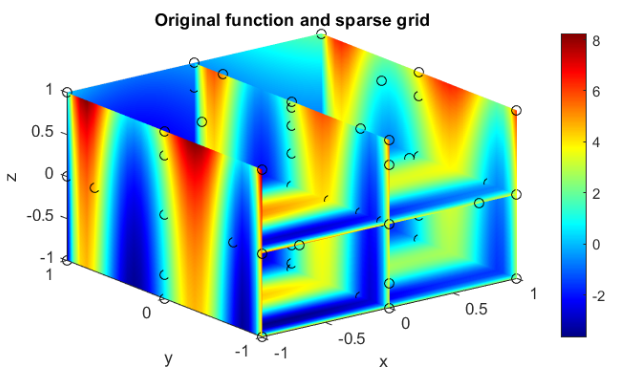}
\caption{Plot of the original function and the training set (circles) used to construct the PGD models.}\label{s2PGD_ex1_grid}
\end{center}
\end{figure}

Moreover, the sampling for the training set is created using a sparse grid based on the Smolyak quadrature rule \cite{Key_Smolyak,PhD_sparsegrids} of level 3 based on the Clenshaw-Curtis univariate quadrature rule. Therefore, only these points are used to construct the model either using the $s$-PGD or the $s^2$-PGD methodology. In figure \ref{s2PGD_ex1_grid}, the mesh used for the training set is shown.

On the other hand, a testing set of 27000 untrained points is available to compare the results between techniques when predicting unseen scenarios. This second set will be used to study the predictive ability of both models once they are finally constructed.

The conditions to employ the $s^2$-PGD in this example are the following. A basis reaching eighth-degree polynomials is chosen for the sparse dimension. Moreover, a standard MAS-based $s$-PGD is used, reaching 4th degree polynomials along the non-sparse dimensions. 

\begin{figure}
\begin{center}
\includegraphics[width=0.7\linewidth]{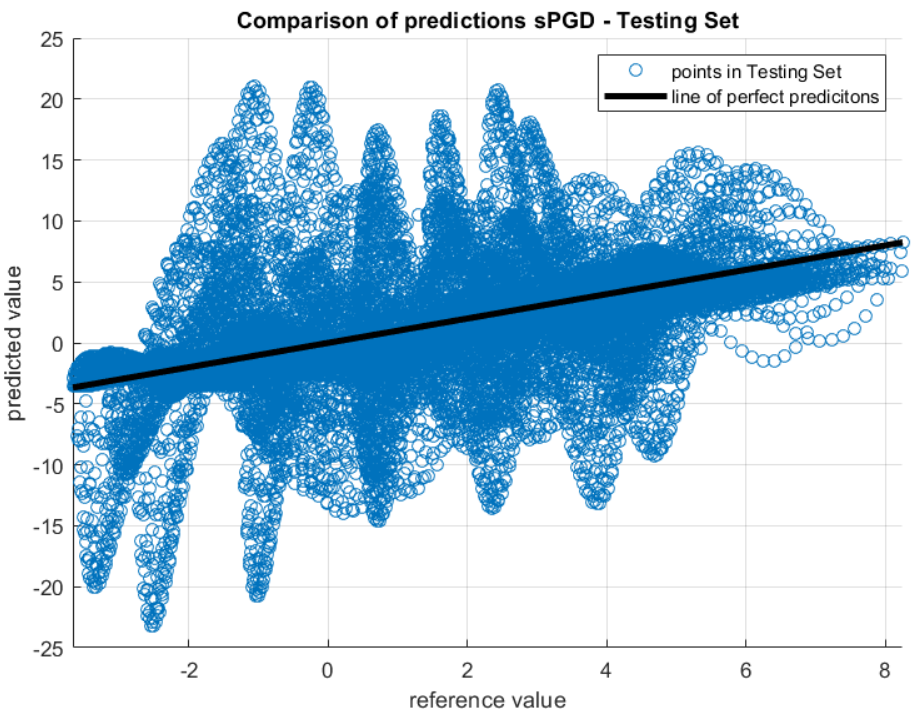}
\caption{Problem defined in Eq. \eqref{eqn:s2PGD_a}: Comparison of predicted $s$-PGD values with the reference ones in the testing set (the black line represents a perfect prediction)}\label{s2PGD_ex1_OLS}
\end{center}
\end{figure}

In Figure \ref{s2PGD_ex1_OLS}, the results of the standard $s$-PGD are shown. In this case, we can see that the predictions are bad because this methodology completely fails in finding this type of sparse solutions. This is one of the problems that the $s$-PGD is facing and we propose to solve with the $s^2$-PGD. 

In addition, if we observe the $s$-PGD solution we can see that all the possible elements are nonzero, so it fails in identifying the sparsity. To detect sparsity, three simulations of the $s^2$-PGD are carried out, penalizing a different dimension at each iteration. Consequently, the model with best predictive ability (out of the training set) will be the selected one. As expected, the chosen model is the one obtained when penalizing the $x_2$ dimension.

\begin{figure}
\begin{center}
\includegraphics[width=0.7\linewidth]{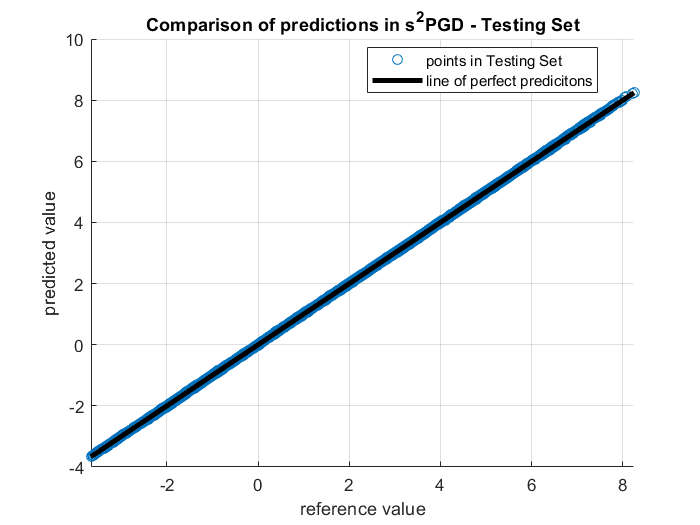}
\caption{Problem defined in Eq. \eqref{eqn:s2PGD_a}: Comparison of predicted $s^2$-PGD values with the reference ones in the testing set (the black line represents a perfect prediction)}\label{s2PGD_ex1}
\end{center}
\end{figure}

In Figure \ref{s2PGD_ex1}, the results of the $s^2$-PGD are presented. As we can observe, predictions are almost perfect. If we examine the solution, we can see that the model is correctly identified using four modes, that is, four sums of the PGD decomposition.

The errors concerning the $s$-PGD and the $s^2$-PGD solutions are respectively $\text{err}_{pgd} = 141$ \% and $\text{err}_{s^2pgd} = 0.56$ \%.

\subsubsection{A second example involving more dimensions}

In this case we consider the approximations problem of function
\begin{multline}\label{eqn:s2PGD_b}
f(x_1, x_2, x_3, x_4, x_5) = 
\big[ T_5(x_1) + 2T_1(x_1) \big] 
\big[ T_2(x_2) + 2T_4(x_2) \big] \\
\Big[ 
\big( \sin(2x_3) - 3.14 \big) \log(3x_4 + 1.5) \cos(x_5) 
+ 
\exp(x_4) \cosh(x_3) \sinh(x_5)
\Big]
\end{multline}
by using a Chebyshev approximation basis for the one-dimensional functions involved in the PGD constructor.

The above function is intended to be reconstructed in the domain $\Omega = [-1, 1]^5$. The sampling for the training set contains 290 points. In addition, the Latin hypercube sampling is used to generate this random set of data.

On the other hand, a testing set of 2000 untrained points is available to compare the results between techniques when predicting unseen scenarios. As in the previous examples, this second set will be used to study the predictive ability of both models once they are finally constructed.

Concerning the $s^2$-PGD a  basis reaching sixth-degree polynomials is chosen for the sparse dimensions.
Moreover, a standard MAS is used, up-to 4th degree polynomials, in the non-sparse dimensions. 

\begin{figure}
\begin{center}
\includegraphics[width=0.7\linewidth]{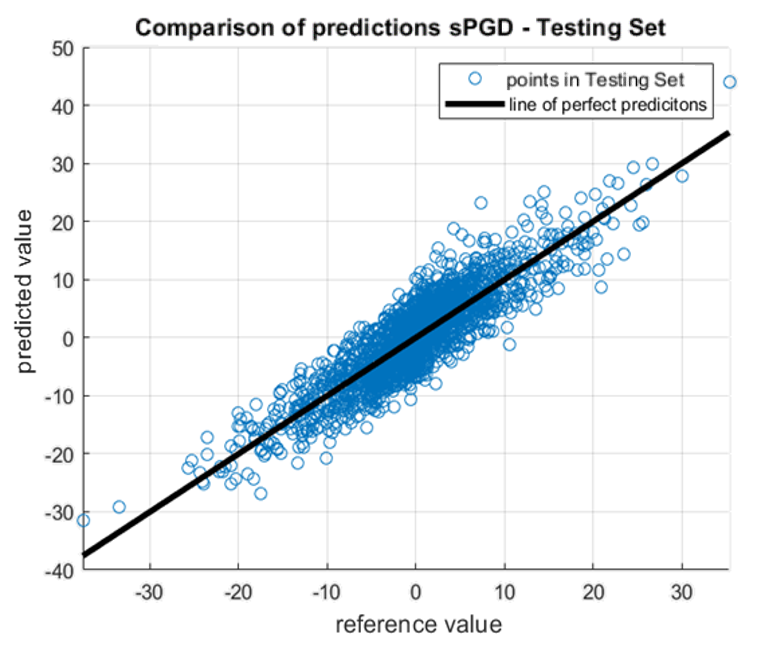}
\caption{Problem defined in Eq. \eqref{eqn:s2PGD_b}: Comparison of predicted $s$-PGD values with the reference ones in the testing set (the black line represents a perfect prediction)}\label{s2PGD_ex2_OLS}
\end{center}
\end{figure}

In Figure \ref{s2PGD_ex2_OLS}, the results of the standard $s$-PGD are shown. In this case, we can see that the predictions are bad. This is due to the wrong identification of the non-zero elements in the separated representation, which causes overfitting problems.

To detect sparsity, five different simulations of the $s^2$-PGD are carried out, penalizing one different dimension each time. Consequently, the model with best predictive ability (out of the training set) will be the selected one.
As expected, the chosen model is the one obtained when penalizing the $x_1$
dimension. The reason is that in this case, we observe that the correct non-zero terms for $x_1$ and $x_2$ are identified just penalizing $x_1$.

\begin{figure}
\begin{center}
\includegraphics[width=0.7\linewidth]{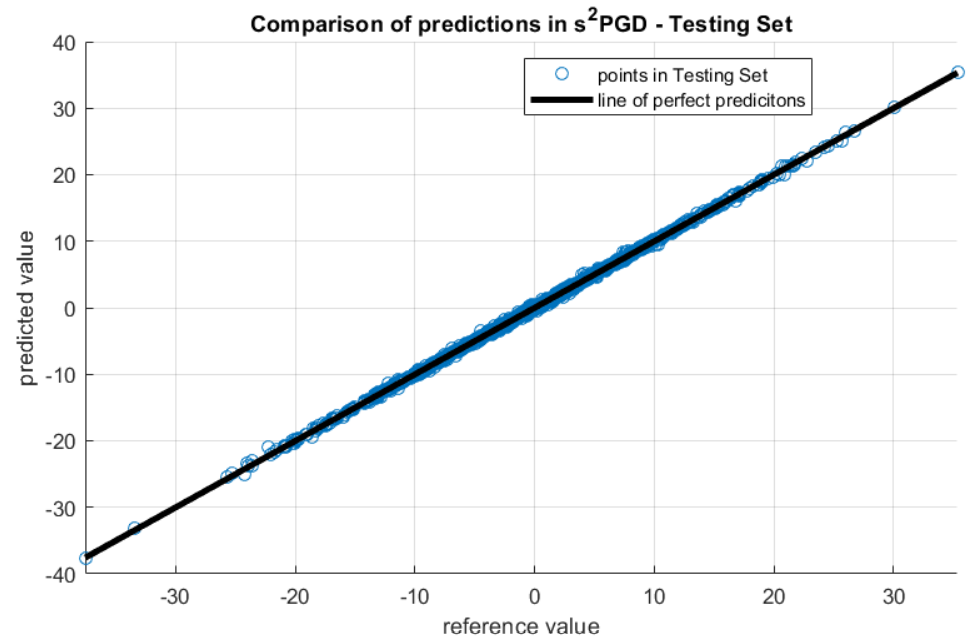}
\caption{Problem defined in Eq. \eqref{eqn:s2PGD_b}: Comparison of predicted $s^2$-PGD values with the reference ones in the testing set  (the black line represents a perfect prediction)}\label{s2PGD_ex2}
\end{center}
\end{figure}

In Figure \ref{s2PGD_ex2}, the results of the $s^2$-PGD are presented. An exellent agreement between the real function and the proposed approach is observed. Furthermore, if we examine the $s^2$-PGD solution, we can see that the model have correctly identified the non-zero elements. In addition, this PGD solution needed 104 modes, that is, 104 sums of the PGD decomposition, solution that can be re-compacted by invoking again the PGD  \cite{BookPGD_Paco}.

Finally, the errors concerning the $s$-PGD and the $s^2$-PGD solutions are respectively $\text{err}_{pgd} = 46.39$ \% and $\text{err}_{s^2pgd} = 2.4$ \%.

\subsection{ANOVA-PGD numerical results}
\label{anov}

ANOVA-PGD regression consists of applying regression techniques (such as standard interpolation, $s$-PGD, $rs$-PGD or $s^2$-PGD) separately to the different terms (or groups of terms) in the ANOVA decomposition. This strategy suggests the MAS since it enforces some simplicity in the first modes, even if here richer approximations can be envisaged, but it also provides other benefits through the orthogonality of the decomposition and the opportunity to work in a low dimension setting, as previously expossed.

Here, we consider the numerical test related to the 2D function
\begin{equation}\label{eqn:anova}
f(x,y)=-2\cos(3x^{1.75})+10\log(y-0.6)^4+6\cos(x)(y-0.3y^2),
\end{equation}
that perfectly fits the ANOVA structure, despite the functional complexity of the terms involving the coordinates $x$ and $y$, $2\cos(3x^{1.75})$ and $10\log(y-0.6)^4$ respectively, and the one coupling both coordinates, $6\cos(x)(y-0.3y^2)$.

When considering the ANOVA-based sampling consisting of the center point of the parametric domain acting as the anchor $\vec{c}=(x_c,y_c)$, 10 additional points in the first dimension (of the form $(x,y_c)$) and 10 additional points in the second dimension (of the form $(x_c,y)$), functions $f_x(x)$ and $f_y(y)$ were calculated with a cubic spline interpolation. Then, a  standard 2D nonlinear regression using basis functions of the form $(x-x_c)^m(y-y_c)^n,\ m,n\geq 1$ (due to the low dimensionality of the treated problem the employ of separated representations is not needed) was employed for calculating the term $f_{x,y}(x,y)$ using 4 sample points. 

\begin{figure}
\begin{center}
\includegraphics[width=\linewidth]{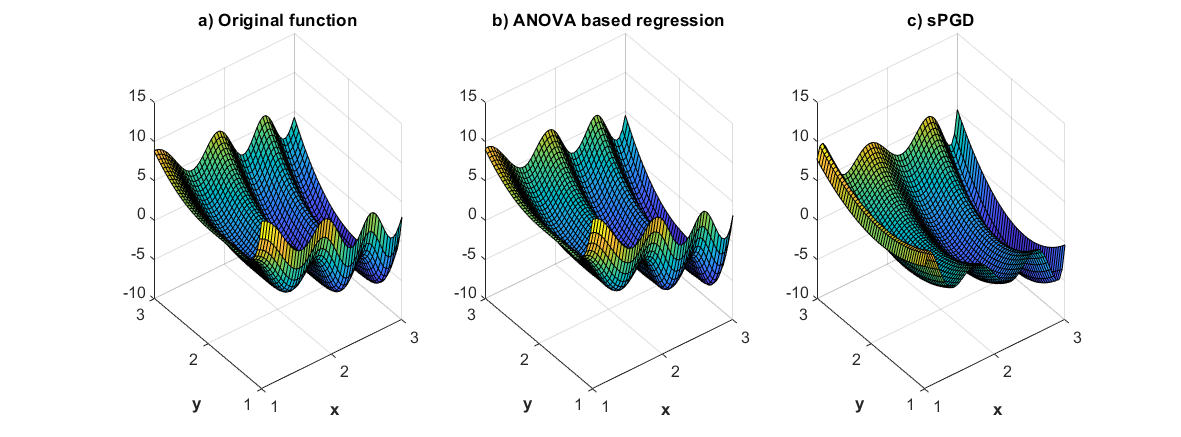}
\caption{Comparing $s$-PGD and ANOVA-PGD regressions}\label{anova_ex1}
\end{center}
\end{figure}

\begin{figure}
\begin{center}
\includegraphics[width=0.75\linewidth]{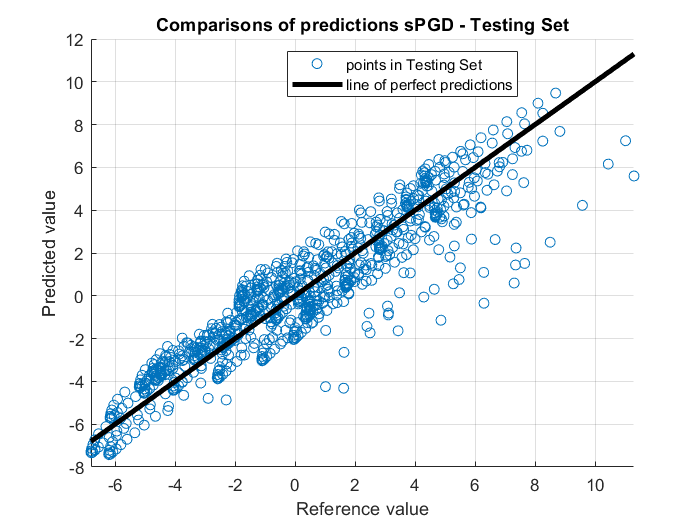}
\caption{Problem defined in Eq. \eqref{eqn:anova}: Comparison of predicted $s$-PGD values with the reference ones in the testing set  (the black line represents a perfect prediction) }\label{anova_ex1_qoi}
\end{center}
\end{figure}

\begin{figure}
\begin{center}
\includegraphics[width=0.75\linewidth]{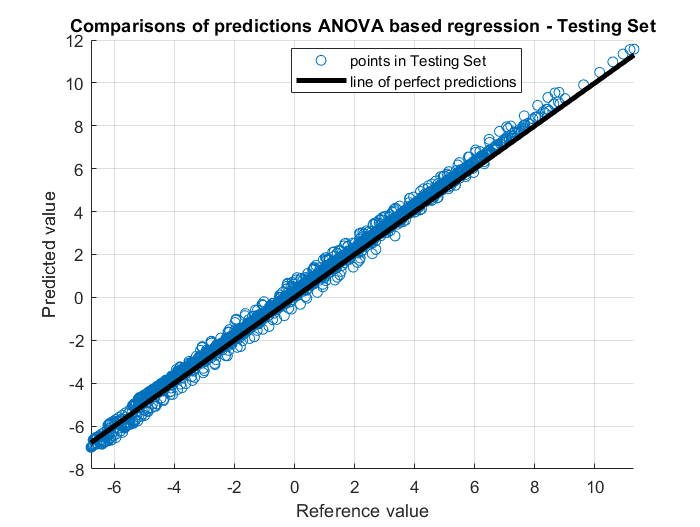}
\caption{Problem defined in Eq. \eqref{eqn:anova}: Comparison of predicted ANOVA-PGD values with the reference ones in the testing set  (the black line represents a perfect prediction) }\label{anova_ex1_qoi_spgd}
\end{center}
\end{figure}

The constructed solution is depicted in Fig. \ref{anova_ex1} where it is compared with the exact solution as well as with the solution obtained by using the standard $s$-PGD (with a Latin Hypercube Sampling containing 25 points), while Figs. \ref{anova_ex1_qoi} and \ref{anova_ex1_qoi_spgd} compare the predictions and the reference values. From all these results, excellent performances of the ANOVA-based regression can be stressed. 
 
\section{Conclusions}
\label{conclusions}

In this paper, three different data-driven regression techniques are introduced, the first two, the so-called $rs$-PGD and $s^2$-PGD, that consist of a regularization of the usual sparse PGD, and the third, that combines analysis of variance features with sparse separated representations.  It has been shown and discussed, through different examples, how they can improve significantly the existing sparse $s$-PGD performance, reducing overfitting and achieving great explanatory predictive capabilities when dealing with unseen scenarios.

Furthermore, the  $s^2$-PGD can be employed to sparse identification and variable selection when the $s$-PGD  fails. The comparison of Figures \ref{s2PGD_ex1_OLS} and \ref{s2PGD_ex1} is an example of the substantial improvements under this rationale.

In addition, the suitability of the $s$-PGD to deal with the challenging scenarios concerning the low-data regime context and high-dimensional parametric functions was previously proved in \cite{sPGD_Ruben} and \cite{PhD_Ruben}. Therefore, the improvements carried out by these new techniques opens the door to construct better high-performance ROMs in this difficult context. Moreover, this is really appealing because of the increasing industrial interest of obtaining accurate models under these circumstances.


Our works in progress address specific industrial applications where the use of these techniques can be competitively advantageous. In addition, other penalties are being studied for its use in specific frameworks as well as different sampling strategies when they can be controlled, to maximize the ROM performance.

\section*{Compliance with Ethical Standards} The authors declare that they have no conflict of interest.


\begin{thebibliography}{10}

\bibitem{PhD_Clara}
C.~Argerich.
\newblock {\em Study and development of new acoustic technologies for nacelle
  products}.
\newblock PhD thesis, Universitat Politecnica de Catalunya, Jun 2020.

\bibitem{PhD_sparsegrids}
Karim Beddek.
\newblock {\em Propagation d'incertitudes dans les mod{\`e}les {\'e}l{\'e}ments
  finis en {\'e}lectromagn{\'e}tisme : application au contr{\^o}le non
  destructif par courants de Foucault}.
\newblock PhD thesis, Ecole doctorale Sciences pour l'Ingenieur (Lille) - L2EP,
  2012.
\newblock Th{\`e}se de doctorat dirig{\'e}e par Cl{\'e}net, St{\'e}phaneLe
  Menach, Yvonnick et Moreau, Olivier G{\'e}nie {\'e}lectrique Lille 1 2012.

\bibitem{SSL}
Domenico Borzacchiello, Jose~Vicente Aguado, and Francisco Chinesta.
\newblock Non-intrusive sparse subspace learning for parametrized problems.
\newblock {\em Archives of Computational Methods in Engineering},
  26(2):303--326, 2019.

\bibitem{Kutz_data_driven_2019}
S.~L. Brunton and J.~N. Kutz.
\newblock {\em Data-driven science and engineering: Machine learning, dynamical
  systems, and control.}
\newblock Cambridge University Press, 2019.

\bibitem{Lorentz_1}
Steven Brunton, Joshua Proctor, and J.~Kutz.
\newblock Discovering governing equations from data: Sparse identification of
  nonlinear dynamical systems.
\newblock {\em Proceedings of the National Academy of Sciences},
  113:3932–3937, 09 2015.

\bibitem{brunton2016discovering}
Steven~L Brunton, Joshua~L Proctor, and J~Nathan Kutz.
\newblock Discovering governing equations from data by sparse identification of
  nonlinear dynamical systems.
\newblock {\em Proceedings of the national academy of sciences},
  113(15):3932--3937, 2016.

\bibitem{MOR_Paco}
F.~Chinesta, A.~Huerta, G.~Rozza, and K.~Willcox.
\newblock {\em Encyclopedia of Computational Mechanics}, chapter Model Order
  Reduction.
\newblock John Wiley \& Sons, Ltd, 2015.

\bibitem{chinesta2020virtual}
Francisco Chinesta, Elias Cueto, Emmanuelle Abisset-Chavanne, Jean~Louis Duval,
  and Fouad El~Khaldi.
\newblock Virtual, digital and hybrid twins: a new paradigm in data-based
  engineering and engineered data.
\newblock {\em Archives of computational methods in engineering},
  27(1):105--134, 2020.

\bibitem{BookPGD_Paco}
Francisco Chinesta, Roland Keunings, and Adrien Leygue.
\newblock {\em The Proper Generalized Decomposition for Advanced Numerical
  Simulations: A Primer}.
\newblock Springer Publishing Company, Incorporated, 2013.

\bibitem{BookPGD_cueto}
Elias Cueto, David Gonzalez, and Icar Alfaro.
\newblock {\em Proper Generalized Decompositions: An Introduction to Computer
  Implementation with Matlab}.
\newblock Springer Publishing Company, Incorporated, 1st edition, 2016.

\bibitem{SUR1}
A.I.J. Forrester, A.~Sobester, and A.J. Keane.
\newblock {\em Engineering Design via Surrogate Modelling: A Practical Guide}.
\newblock John Wiley \& Sons, Ltd, 2008.

\bibitem{Key_Stats}
T.~Hastie, R.~Tibshirani, and J.~H. Friedman.
\newblock {\em The elements of statistical learning: data mining, inference,
  and prediction}.
\newblock New York: Springer., 2009.

\bibitem{hernandez2020deep}
Quercus Hernandez, Alberto Badias, David Gonzalez, Francisco Chinesta, and
  Elias Cueto.
\newblock Deep learning of thermodynamics-aware reduced-order models from data.
\newblock {\em arXiv preprint arXiv:2007.03758}, 2020.

\bibitem{HERNANDEZ2021109950}
Quercus Hern{\'a}ndez, Alberto Bad{\'\i}as, David Gonz{\'a}lez, Francisco
  Chinesta, and El{\'\i}as Cueto.
\newblock Structure-preserving neural networks.
\newblock {\em Journal of Computational Physics}, 426:109950, 2021.

\bibitem{ibanez2019some}
R~Ibanez, E~Abisset-Chavanne, E~Cueto, A~Ammar, J-L Duval, and F~Chinesta.
\newblock Some applications of compressed sensing in computational mechanics:
  model order reduction, manifold learning, data-driven applications and
  nonlinear dimensionality reduction.
\newblock {\em Computational Mechanics}, 64(5):1259--1271, 2019.

\bibitem{PhD_Ruben}
Ruben Ibanez.
\newblock {\em {Advanced physics-based and data-driven strategies}}.
\newblock Theses, {{\'E}cole centrale de Nantes ; Universitat polit{\'e}cnica
  de Catalunya}, September 2019.

\bibitem{ibanez2018multidimensional}
Rub{\'e}n Ibanez, Emmanuelle Abisset-Chavanne, Amine Ammar, David Gonz{\'a}lez,
  El{\'\i}as Cueto, Antonio Huerta, Jean~Louis Duval, and Francisco Chinesta.
\newblock A multidimensional data-driven sparse identification technique: the
  sparse proper generalized decomposition.
\newblock {\em Complexity}, 2018.

\bibitem{sPGD_Ruben}
Rub{\'e}n Ibanez~Pinillo, Emmanuelle Abisset-Chavanne, Amine Ammar, David
  Gonz{\'a}lez, Elias Cueto, Antonio Huerta, Jean Louis~Duval, and Francisco
  Chinesta.
\newblock A multidimensional data-driven sparse identification technique: The
  sparse proper generalized decomposition.
\newblock {\em Complexity}, 2018:1--11, 11 2018.

\bibitem{Key_Smolyak}
Vesa Kaarnioja.
\newblock {\em Smolyak Quadrature}.
\newblock mathesis, University of Helsinki, 2013.

\bibitem{laughlin2000cover}
Robert~B Laughlin and David Pines.
\newblock The theory of everything.
\newblock {\em Proceedings of the national academy of sciences of the United
  States of America}, 97(1):28, 2000.

\bibitem{Lorentz_2}
Edward~N Lorenz.
\newblock Deterministic nonperiodic flow.
\newblock {\em Journal of the Atmospheric Sciences}, 20:130--141, 1963.

\bibitem{LY01}
H.V. Ly and H.T. Tran.
\newblock Modeling and control of physical processes using proper orthogonal
  decomposition.
\newblock {\em Journal of Mathematical and Computer Modeling},
  33(1-3):223--236, 2001.

\bibitem{moya2020physically}
Beatriz Moya, Iciar Alfaro, David Gonzalez, Francisco Chinesta, and El{\'\i}as
  Cueto.
\newblock Physically sound, self-learning digital twins for sloshing fluids.
\newblock {\em PLoS One}, 15(6):e0234569, 2020.

\bibitem{moya2020digital}
Beatriz Moya, Alberto Bad{\'\i}as, Ic{\'\i}ar Alfaro, Francisco Chinesta, and
  El{\'\i}as Cueto.
\newblock Digital twins that learn and correct themselves.
\newblock {\em International Journal for Numerical Methods in Engineering},
  2020.

\bibitem{moya2019learning}
Beatriz Moya, David Gonz{\'a}lez, Ic{\'\i}ar Alfaro, Francisco Chinesta, and
  E~Cueto.
\newblock Learning slosh dynamics by means of data.
\newblock {\em Computational Mechanics}, 64(2):511--523, 2019.

\bibitem{SUR2}
P.~P.~Jiang, Q.~Zhou, and X.~Shao.
\newblock {\em Surrogate Model-Based Engineering Design and Optimization}.
\newblock Springer, 2020.

\bibitem{KRI}
A.~Papritz and A.~Stein.
\newblock {\em Surrogate Model-Based Engineering Design and Optimization}.
\newblock Stein A., Van der Meer F., Gorte B. (eds) Spatial Statistics for
  Remote Sensing. Remote Sensing and Digital Image Processing, vol 1. Springer,
  Dordrecht, 1999.

\bibitem{sancarlos2020rom}
Abel Sancarlos, Morgan Cameron, Andreas Abel, Elias Cueto, Jean-Louis Duval,
  and Francisco Chinesta.
\newblock From rom of electrochemistry to ai-based battery digital and hybrid
  twin.
\newblock {\em Archives of Computational Methods in Engineering}, pages 1--37,
  2020.

\bibitem{Abel_Motors1}
Abel Sancarlos, Elias Cueto, Francisco Chinesta, and JL~Duval.
\newblock A novel sparse reduced order formulation for modeling electromagnetic
  forces in electric motors.
\newblock {\em SN Applied Sciences}, 2021.

\bibitem{Abel_PGDTFM}
Abel Sancarlos, Manuel Pineda, Ruben Puche, Angel Sapena, Martin Riera,
  J.~Martinez, Juan Perez, and Jose Roger.
\newblock Application of the parametric proper generalized decomposition to the
  frequency-dependent calculation of the impedance of an ac line with
  rectangular conductors.
\newblock {\em Open Physics}, 15, 12 2017.

\bibitem{shiffrin2020brain}
Richard~M Shiffrin, Danielle~S Bassett, Nikolaus Kriegeskorte, and Joshua~B
  Tenenbaum.
\newblock The brain produces mind by modeling.
\newblock {\em Proceedings of the National Academy of Sciences},
  117(47):29299--29301, 2020.

\bibitem{udrescu2020ai}
Silviu-Marian Udrescu, Andrew Tan, Jiahai Feng, Orisvaldo Neto, Tailin Wu, and
  Max Tegmark.
\newblock Ai feynman 2.0: Pareto-optimal symbolic regression exploiting graph
  modularity.
\newblock {\em arXiv preprint arXiv:2006.10782}, 2020.

\end{thebibliography}
\end{document}